\documentclass[showpacs,aps,prd,reprint,groupedaddress,superscriptaddress,preprintnumbers,floatfix,a4paper,nofootinbib,amsmath,amssymb]{revtex4-1}
\usepackage{slashed}
\usepackage{graphicx}
\usepackage{booktabs}
\usepackage[hidelinks]{hyperref}
\usepackage{color}
\hypersetup{
    colorlinks,
    linkcolor={blue},
    citecolor={blue},
    urlcolor={blue}
}
\usepackage{multirow} % Create tabular cells spanning multiple rows
\usepackage{amssymb,amsmath,amsfonts}
\usepackage[utf8]{inputenc}
\usepackage{nicefrac}
\usepackage{braket}
\usepackage{slashed}
\usepackage{soul}

\def\chitop{\chi_\mathrm{top}}
\def\Ndof{N_{\mathrm{dof}}}
\def\DD{\mathcal{D}}
\def\Eq#1{Eq.~(\ref{#1})}
\def\plaqs{\begin{picture}(4,4) \multiput(0,0)(0,4){2}{\line(1,0){4}}
		\multiput(0,0)(4,0){2}{\line(0,1){4}}\end{picture}}
\def\plaq{\begin{picture}(6,6) \thicklines\multiput(0,0)(0,6){2}{\line(1,0){6}}
		\multiput(0,0)(6,0){2}{\line(0,1){6}}\end{picture}}
\def\betamin{\beta_{\mathrm{min}}}
\def\betamid{\beta_{\mathrm{mid}}}
\def\betamax{\beta_{\mathrm{max}}}
\def\Smin{S_{\mathrm{min}}}
\def\Smid{S_{\mathrm{mid}}}
\def\Smax{S_{\mathrm{max}}}
\def\Tc{T_{\mathrm{c}}}
\def\dd{\mathrm{d}}

\let\originalleft\left
\let\originalright\right
\renewcommand{\left}{\mathopen{}\mathclose\bgroup\originalleft}
\renewcommand{\right}{\aftergroup\egroup\originalright}

\begin{document}
	
\title{\boldmath		Multicanonical reweighting for the QCD topological susceptibility }
	
\author{P. Thomas Jahn}
\email{tjahn@theorie.ikp.physik.tu-darmstadt.de}
\author{Parikshit M. Junnarkar}
\email{parikshit@theorie.ikp.physik.tu-darmstadt.de}
\author{Guy D. Moore}
\email{guymoore@theorie.ikp.physik.tu-darmstadt.de}
\affiliation{Institut f\"ur Kernphysik (Theoriezentrum), Technische Universit\"at Darmstadt,\\ Schlossgartenstra{\ss}e 2, D-64289 Darmstadt, Germany}
\author{Daniel Robaina}
\email{daniel.robaina@mpq.mpg.de}
\affiliation{Max-Planck-Institut f\"ur Quantenoptik, Hans-Kopfermann-Stra{\ss}e 1, D-85748 Garching, Germany}

\begin{abstract}
{
We introduce a reweighting technique which allows for a continuous
sampling of temperatures in a single simulation and employ it to
compute the temperature dependence of the QCD topological
susceptibility $\chitop$ at high temperatures.
The method determines the ratio of $\chitop$ between any two
temperatures within the explored temperature range.
We find that the results from the method agree with our previous
determination and that it is competitive with but not better than
existing methods of determining $\dd\chitop/\dd T$.
The method may also be useful in exploring the temperature dependence
of other thermodynamical observables in QCD in a continuous way.}
\end{abstract}
	
\date{\today}
\maketitle

\section{Introduction}
The axion solution to the strong CP problem, proposed more than four
decades ago~\cite{Peccei:1977hh,Weinberg:1977ma,Wilczek:1977pj},
solves the fine tuning problem of the smallness of the $\theta$
parameter in QCD by introducing a new light ($<\mathrm{meV}$) pseudo
Goldstone boson in the Standard model.  Soon after it was proposed,
it was also realized that long-wavelength axions would be produced
abundantly early in the Big Bang, and can serve as a candidate for
dark matter \cite{Abbott:1982af,Preskill:1982cy}.
These proposals have heightened interest both in searching
experimentally for the axion, and in better understanding its
phenomenology and cosmological history; for a review of these topics
see for instance Ref.~\cite{Irastorza:2018dyq}.
An important input for the cosmological production of axions
and for their contemporary properties is the QCD topological
susceptibility $\chitop(T)$, which determines the axion mass:
\begin{equation}
m^2_{\mathrm{A}}(T) = \frac{\chitop(T)}{f^2_{\mathrm{A}}}.
\end{equation}
Here $m_{\mathrm{A}}(T)$ is the temperature dependent axion mass and $f_{\mathrm{A}}$ is the
axion decay constant.  The low-temperature value of $\chitop$ is well
determined \cite{diCortona:2015ldu,Gorghetto:2018ocs}, but the
high-temperature regime determines axion production efficiency;
it is particularly important to determine $\chitop(T)$ in the
temperature regime from 500 MeV to 1200 MeV \cite{Moore:2017ond}.
In this temperature range, this can only be achieved by
nonperturbative lattice investigations \cite{Borsanyi:2015cka}.
The most straightforward lattice methods, based on brute-force
sampling of the gauge field configuration space, face difficulties in
this temperature range because topologically nontrivial gauge field
configurations become very rare, leading to a loss of statistical
power.  New methodologies are needed to overcome this problem.
In recent years one such methodology has been developed \cite{Borsanyi:2016ksw,Frison:2016vuc} which provides access to high temperatures. 
In these calculations, it was shown that the difference of the
expectation of the QCD action in two topological sectors can provide a
determination of $\dd(\ln\chitop)/\dd(\ln T)$, which can be integrated
to provide the temperature dependence of $\chitop$.

Alternatively, one can approach the problem at a fixed, high
temperature by reweighting between topological sectors.  A first
attempt, based on a fixed guess for the reweighting function
\cite{Bonati:2018blm}, explored temperatures around 500 MeV.
Another method, in which the reweighting function is determined
dynamically via an iterative self-consistent technique, was introduced in
\cite{Jahn:2018dke} and further improved in \cite{Jahn:2020oqf}, where
it was shown to be effective in the pure-glue theory up to at least
$7 \, \Tc$.

Our main motivation for this work is to explore whether a
new technique might improve existing methods for determining $\chitop$.
The method introduced here is related to but distinct from the technique of Refs
\cite{Borsanyi:2016ksw,Frison:2016vuc}.  
We train a single Markov-chain Monte-Carlo simulation to explore a wide range of
temperatures in a detailed-balance respecting way by replacing the
weighting function $\exp[-\beta S]$ with $\exp[-W(S)]$, where $W(S)$
is established by an iterative procedure such that the resulting
ensemble can be reweighted to describe any temperature in a relatively
wide range.  We do this separately in the $Q=0$ and $|Q|=1$
(non-topological and instanton-number $\pm 1$) sectors, and combine
with a determination of $\chitop(T)$ at one temperature to determine
$\chitop$ across the full accessible temperature range.
We show that the approach has a similar efficiency to
\cite{Borsanyi:2016ksw}, with both (slight) advantages and
disadvantages relative to their approach.
The technique can be extended to include fermions if the line of
constant physics (lattice $m(\beta)$) is known.  However we find that
our approach developed in Ref.~\cite{Jahn:2020oqf} seems to afford
better numerical efficiency.

In the next section we review the definition of topological
susceptibility and lay the groundwork for our technique.  Section
\ref{sec:reweight} introduces our technique to carry out a
detailed-balance preserving Markov chain over a range of
temperatures.  Then Section \ref{sec:results} presents our results and
Section \ref{sec:discussion} closes with a discussion.

\section{Topological Susceptibility}
\label{sec:chitop}

The topological susceptibility is defined as
\begin{equation}
\label{chi2}
\chitop(T) = \frac{ T}{V} \; \langle Q^2 \rangle 
   = \frac{T}{V} \int \dd^4 x \; \dd^4 y \;	\langle q(x) q(y) \rangle, 
\end{equation}
where $Q = \int \dd^4 x \: q(x)$ and $q$ is the topological charge
density, $V$ the spatial volume and $T$ the temperature.
In the continuum $Q$ always takes an integer value while on the
lattice one requires a refined definition of $Q$.
Considering the continuum integer topological charge $Q$  one can
write the partition function as a sum over topological sectors:
\begin{equation}
	\label{Zsum}
	Z = \int \DD A e^{-\beta S}
	= \sum_{N \in \mathcal{Z}} \int \DD A e^{-\beta S} \delta(Q-N)
	\equiv \sum_{N} Z_N \,.
\end{equation}
The susceptibility is then given by : 
\begin{eqnarray}
	\label{chi3}
	\chitop(T) &=& \frac{T}{V}
	\frac{\int \DD A e^{-\beta S} Q^2}{\int \DD A e^{-\beta S}}
	\nonumber \\ & = &
	\frac{T}{V} \frac{\sum_N N^2 Z_N}{\sum_N Z_N} \,.
\end{eqnarray}
At low temperatures and/or large volumes, this sum will have important
contributions from many $N$ values.  But at a sufficiently high
temperature, such that $V/T \; \chitop(T) \ll 1$, the numerator is dominated
by $N=1$ and $N=-1$ and the denominator is dominated by $N=0$.
Renaming $Z_1 + Z_{-1} \to Z_1$ (the part of the partition function
where $Q^2 = 1$, that is, $Q = \pm 1$) and considering a finite volume we find
\begin{align}
	\label{chi4}
	\chitop(T) \simeq \frac{T}{ V} \frac{Z_1}{Z_0}
	& = \frac{1}{a^4 N_x N_y N_z N_\tau} \frac{Z_1}{Z_0} \nonumber \\
	\chitop(\beta) a^4(\beta) & = \frac{1}{V_L}  \frac{Z_1}{Z_0}.
\end{align}
Here we have re-expressed the temperature and volume as they would
appear in a lattice calculation, with lattice spacing $a$ and lattice
extents $(N_\tau,N_x,N_y,N_z)$ in the temporal and the three space
directions, and $V_L = N_\tau N_x N_y N_z$ the number of lattice
sites.  The last line emphasizes that the lattice spacing is
a function of the lattice gauge coupling
$\beta = 6/g^2_{\mathrm{latt}}$.  The nontrivial relation between
these two quantities is determined by a scale setting measurement.

In this work, we will present a method which determines the ratio of
susceptibilities at two specified temperatures.  With a fixed lattice
extent in the spatial and temporal directions, the two temperatures will
correspond to two gauge couplings $\beta_\mathrm h$ and $\beta_\mathrm c$ (h for hot
and c for cold) which will
in turn correspond to two different lattice spacings $a(\beta_\mathrm h)$ and
$a(\beta_\mathrm c)$.
The ratio of susceptibilities at two temperatures will then be given as:
\begin{equation}
	\frac{\chitop(\beta_\mathrm h) a^4(\beta_\mathrm h)} {\chitop(\beta_\mathrm c) a^4(\beta_\mathrm c) } =
	\frac{Z_1(\beta_\mathrm h)}{Z_1(\beta_\mathrm c)} \;
	\frac{Z_0(\beta_\mathrm c)}{Z_0(\beta_\mathrm h)}.
	\label{ratios}
\end{equation}
Our method will also allow for the determination of the above ratio for
\textit{any} pair of temperatures within the prespecified range.
We will do so by computing the ratio of partition
functions using two Monte-Carlo simulations: one that works within the
$Q=1$ topological sector and one which works within the $Q=0$ sector,
but with each simulation exploring the full range of $\beta$ values,
as we describe in the next section.

\section{Temperature reweighting}
\label{sec:reweight}

\begin{figure}
	\centering
	\includegraphics[width=\linewidth]{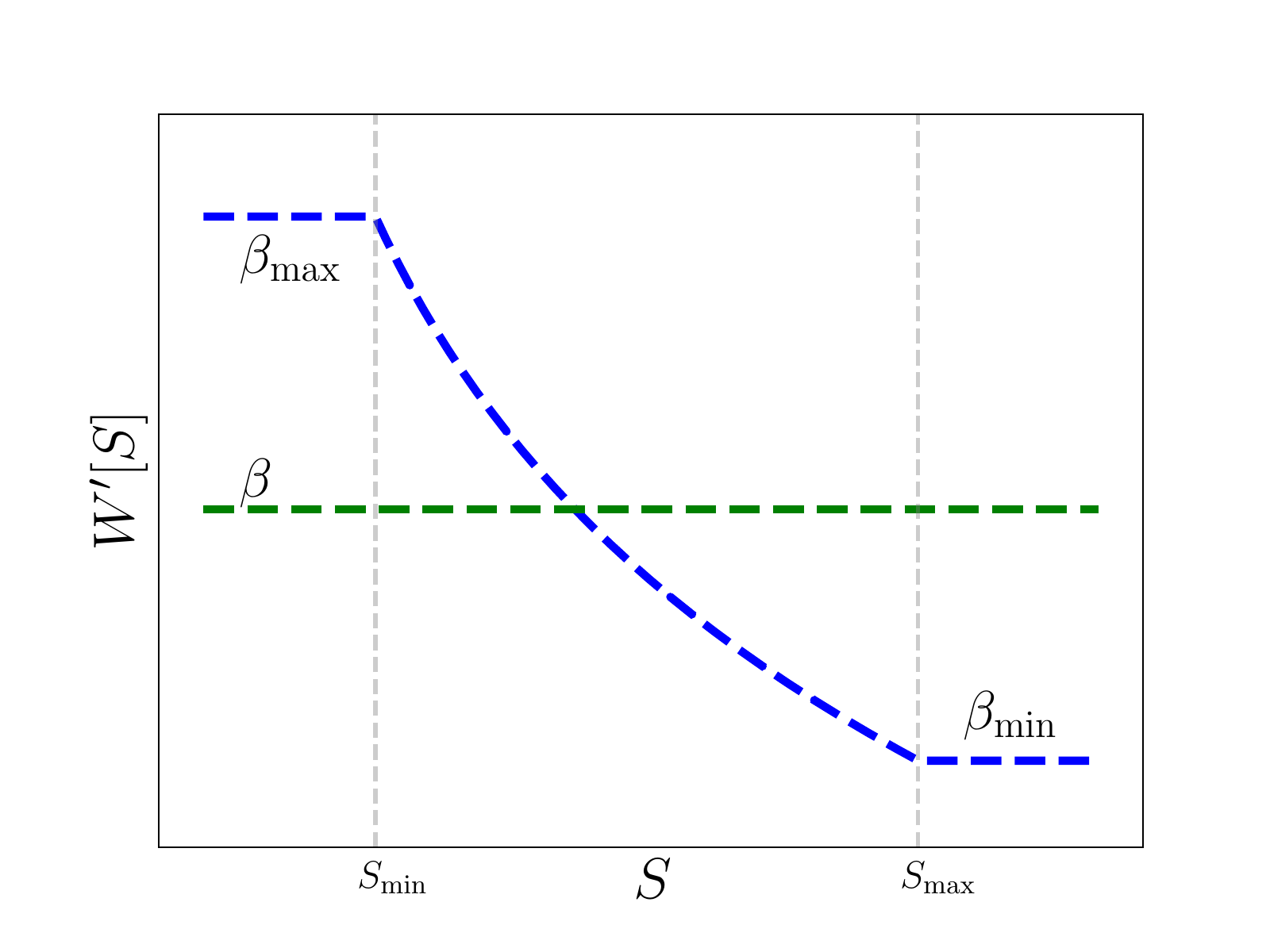}
	\caption{A cartoon of the Monte-Carlo sampling of $W[S]$.
          The green line represents a $W$ choice which will sample one
          $\beta$ value, while the blue line will sample all
          $\beta$ values between $\betamax$ and $\betamin$.}
	\label{fig:cartoon}
\end{figure}
In a given Monte-Carlo simulation, we intend to sample a range of
temperatures continuously over a prespecified range in pure glue QCD
regularized on a finite space-time lattice.
We will accomplish this using a reweighting method very similar to the
old proposal of Berg and Neuhaus \cite{Berg:1991cf}.  In this section
we first show how a reweighted Monte-Carlo can be used to determine
the susceptibility; then we show how the reweighted Monte-Carlo can be
carried out; and finally we show how to determine the reweighting
function itself.  In practice, the numerical work proceeds in exactly
the opposite order.

\subsection{Susceptibility using reweighting}
\label{sec:suscept}

In the standard  Monte-Carlo simulation one evaluates the  partition
function  $\mathcal{Z}(\beta)$ shown below at a particular gauge
coupling $\beta$ by generating gauge configurations with the
probability distribution $\dd P[U]$ as 
\begin{equation}
	\label{usualMC}
	 \mathcal{Z}(\beta) =\int \DD U \; e^{-\beta S} , \quad	\mathrm{d}P[U]= \frac{\mathrm{d}U \; e^{-\beta S}}{\mathcal{Z}(\beta)} .
\end{equation}
\iffalse
For a general case of a  reweighting, one introduces a weight function $W[\xi]$ such that  
\begin{equation}
	\label{ourway1}
	\mathcal{Z}(\beta)= \int \DD U e^{-W[\xi]}  \; e^{+W[\xi] - \beta S}.
\end{equation}
Now in order to sample such a  reweighted partition function with Monte-Carlo one generates gauge configurations with the following  modified distribution d$P[U]$,
\begin{equation}
	\label{propdis}
	\mathrm{d}P[U]= \frac{\mathrm{d}U \; e^{+W[\xi] - \beta S}.}{\int \DD U \; e^{+W[\xi] - \beta S}} .
\end{equation}
In the reweighting function $W[\xi]$, $\xi$ is appropriately chosen reweighting variable which enables the Monte-Carlo to sample the parameter $\xi$.
As was shown in \cite{Jahn:2018dke}, with $\xi=Q$, where $Q$ is improved topological charge, the algorithm's ability to sample topology was enhanced in a high temperature regime where the statistical sampling of the topology is severely restricted with the usual Monte-Carlo method.
\fi
Such a Monte-Carlo simulates a single temperature which is controlled
by the gauge coupling $\beta$ since the lattice dimensions are
fixed.  In order to simulate a range of $\beta$ values and therefore a
range of temperatures, the sampling weight in the usual
Monte-Carlo in \Eq{usualMC}  is replaced with
\begin{equation}
	\int \DD U \; e^{-\beta S}  \longrightarrow \int \DD U  \; e^{-W[S]}.
\end{equation} 
The weight function $W[S]$ is a general function of the action $S$,
and its derivative can be interpreted, approximately, as the gauge
coupling $\beta$ to use at a given action value:
\begin{equation}
	W^\prime[S] \equiv \frac{\mbox{d} W[S]}{\mbox{d}S} = \beta[S].
	\end{equation}
In this sense, the usual Monte-Carlo has simply $W[S] = \beta S $ with the constant slope simulating the fixed coupling.
We show how to cover a range of temperatures schematically in
Figure~\ref{fig:cartoon}.  To get a fixed temperature one picks
$W'[S] = \beta$ as indicated by the green line in the figure; to cover
a range of temperatures, one instead uses the blue line in the figure,
for which $W'[S]$ varies from $\betamax$ to
$\betamin$ as $S$ changes from $S_{\mathrm{min}}$ to
$S_{\mathrm{max}}$;  here $S_{\mathrm{min}}$ and $S_{\mathrm{max}}$
are the average values of $S$ in simulations with coupling strengths
$\betamax$ and $\betamin$ respectively.  For $S_0$ an action value
between these limiting choices, $W'[S_0]$
is chosen to be the $\beta$ value for which the expectation value of
the action would be $S_0$.
Outside the range of interest, the weight function $W[S]$ then
simulates two different constant gauge couplings $\betamax$ and
$\betamin$.  With this choice of $W[S]$, a single Markov chain
Monte-Carlo will sample the entire range of $S$ values between
$S_{\mathrm{min}}$ and $S_{\mathrm{max}}$ rather uniformly.  The
function $W'[S]$ is not known a priori; we will return to its
determination shortly, but will first discuss how a simulation with
such a weight function can be carried out and used.

Once $W[S]$ is known, a Monte-Carlo simulation with $W[S]$ generates
an ensemble which can be reweighted to determine expectation values at
a given $\beta$ value $\beta_0$ via
\begin{eqnarray}
  \label{eq:rwgtZ}
  \mathcal{Z}(\beta_0) &=&
  \int \DD U e^{-W[S]}  \; e^{+W[S] - \beta_0 S}
  \\ \nonumber
  &\propto& \sum_{i} e^{W[S_i] - \beta_0 S_i} \,,
\end{eqnarray}
where $i$ indexes the sampled configurations.
In practice this partition function is used to determine expectation
values for operators via
\begin{equation}
  \label{expectation}
  \langle \mathcal{O} \rangle = \mathcal{Z}^{-1} \int \DD U
  e^{-\beta_0 S} \mathcal{O} \simeq
  \frac{\sum_i e^{W[S_i]-\beta_0 S_i} \mathcal{O}_i}
       {\sum_i e^{W[S_i]-\beta_0 S_i}} \,.
\end{equation}

By establishing two reweighting functions $W[S]$ and $W_Q[S]$ for the
$Q=0$ and $|Q|=1$ ensembles respectively, we can generate
multi-temperature ensembles, labeled by $i$ and $iQ$, which
respectively sample the $Q=0$ and the $|Q|=1$ ensembles across
temperatures.  The ratio needed in \Eq{ratios}
is then given by
\begin{align} &
	\label{ourway} 
	 \frac{\chitop(\beta_\mathrm h) a^4(\beta_\mathrm h)}{\chitop(\beta_\mathrm c) a^4(\beta_\mathrm c)}
 \\ \nonumber = &
	 \left( \frac{
		\left(\sum_{iQ} e^{W_Q[S_{iQ}]} e^{-\beta_\mathrm h S_{iQ}}\right)
		\left( \sum_{i\vphantom{Q}} e^{W[S_i]} e^{-\beta_\mathrm c S_i} \right) }
	{ \left(\sum_{iQ} e^{W_Q[S_{iQ}]} e^{-\beta_\mathrm c S_{iQ}}\right)
		\left( \sum_{i\vphantom{Q}} e^{W[S_i]} e^{-\beta_\mathrm h S_i} \right) }
	\right) \,.
\end{align}
There are two subtleties associated with this expression.
The first is that \Eq{eq:rwgtZ} only determines the partition function
up to an overall multiplicative factor.  However, this multiplicative
factor cancels between the numerator and denominator expressions
computed from the same sample.  The second subtlety is that the
partition function $\mathcal{Z}$ also has severe lattice-spacing
dependent renormalizations, which we cannot easily compute.
Fortunately, these cancel because each lattice spacing occurs once in
the numerator and once in the denominator in \Eq{ourway}.

\subsection{Update algorithm with $W[S]$}
\label{sec:algorithm}

In this section we explain the algorithm to perform a Monte-Carlo simulation with weight function $e^{-W[S]}$. 
Our aim is to generate a sampling with probability distribution
\begin{equation}
	\mbox{d}P[U] = \frac{e^{-W[S[U]]}}{\int \DD U e^{-W[S[U]]} }
\end{equation}
where $S[U]$ is the standard lattice gauge action and  we assume that $W[S[U]]$ is a known differentiable function of the action $S$.
Simulating such a weight requires a slight modification of the
standard Hybrid Monte-Carlo (HMC) algorithm \cite{Duane:1987de}.%
\footnote{It is also straightforward to use a mixture of heatbath and
  overrelaxation steps.  However this approach does not generalize to
  the unquenched theory, so we concentrate on the HMC approach.}

As in the standard HMC algorithm, we introduce canonical momenta
$\pi_\mu$ for the link variables $U_\mu$, and define a Hamiltonian for
this system as
\begin{eqnarray}
	\mathcal{H}\big(\pi, U \big) &\equiv& \sum_{\mu,x} \frac{1}{2} (\pi_\mu(x))^2
	+ W[S_U] \label{hamiltonian}\\ 
	S_U &=& \sum_{\plaqs} \left( 1 - \frac{1}{3} \mathrm{Tr} \: \plaq \right) \label{action}\\ \nonumber
\end{eqnarray}
where $S_U$ in \Eq{action} is the standard Wilson gauge action written
\textit{without} the gauge coupling $\beta$ prefactor.  The standard
HMC would use the same Hamiltonian but with $W[S_U]$ replaced
by $\beta S_U$, that is, it would use a strictly linear function
for $W[S_U]$.

A single HMC update trajectory consists of the standard steps:
\begin{enumerate}
\item
  Picking a random canonical momentum $\pi_\mu(x)$ from a Gau{\ss}ian
  ensemble  independently for each of the elements of the Lie algebra.
\item
  Solving the following Hamilton equations of motion (shown here
  schematically):
  \begin{align}
    \label{eomU}
    \frac{\dd U}{\dd t} & = -i \pi U \,,
    \\
    \label{eomP}
    \frac{\dd\pi}{\dd t} & = i U^\dagger \frac{\partial W[S_U]}{\partial U}
    = i U^\dagger \frac{\dd W[S_U]}{\dd S_U} \frac{\partial S_U}{\partial U}
  \end{align}
  for a total time $t_0$.
  The derivative with respect to $U_\mu$ is a Lie derivative and in this
  sense these equations are schematic representation.
  Here the time $t$ is a fictitious variable under which the Hamiltonian
  $\mathcal{H}$ is conserved.

  These Hamiltonian equations are discretized using a time-symmetric
  solver such as the leapfrog or Omelyan algorithms.  Under these
  algorithms, one iteratively solves \Eq{eomU} for all link variables
  $U$ at fixed $\pi$, and then solves \Eq{eomP} for all $\pi$
  variables at fixed $U$.  Before applying \Eq{eomP}, we must compute
  $S_U$ and use the (instantaneous) value of $W'[S_U]$ in place of the
  usual factor of $\beta$ at each time step during the update.

  The use of a time-symmetric algorithm is essential, since it ensures
  the property that, if the pair $(U_\mathrm i,\pi_\mathrm i)$ (i for initial) is
  carried to $(U_\mathrm f,\pi_\mathrm f)$ under the update algorithm, then the pair
  $(U_\mathrm f,-\pi_\mathrm f)$ is carried to $(U_\mathrm i,-\pi_\mathrm i)$ up to roundoff error
  effects.  This is sufficient to ensure that the algorithm converges,
  in a Fokker-Planck sense, to the probability distribution
  $\exp[-\mathcal{H}]$ provided that we also include a Metropolis accept/reject step.
\item
  In adding the Metropolis step, the change in the Hamiltonian
  $\Delta \mathcal{H} = \mathcal{H}(U_\mathrm f,P_\mathrm f) - \mathcal{H}(U_\mathrm i,P_\mathrm i)$ is
  compared to a random number drawn uniformly from the interval
  $[0,1]$: $\mathcal{R}[0,1]$.  Whenever
  $e^{\Delta\mathcal{H}} < \mathcal{R}[0,1]$, we accept the change,
  and proceed with $U_\mathrm f$ as our new configuration.
  Otherwise we revert to $U_\mathrm i$, that is, we reject the update.
\end{enumerate}
The only differences with respect to the standard HMC algorithm
\cite{Duane:1987de} are
the use of $W'[S_U]$ the ``instantaneous $\beta$ value'' in place of
$\beta$ in \Eq{eomP} and the use of $W[S_U]$ in place of $\beta S_U$
in the Metropolis accept-reject step.
Both modifications are compatible with the time-symmetry of the
update algorithm and therefore preserve detailed balance.
With these modifications, the HMC algorithm now generates the desired probability distribution.

For the case of a $Q=1$ simulation, an additional accept-reject step
is needed, in which we check to see whether the configuration has
fallen down into the $Q=0$ sector and reject the update if this is the
case.  In practice we can buffer every $N$'th configuration and only
perform this step after every $N$ HMC update steps, reverting to the
last buffered configuration when the check fails.
We define $Q$ as the lattice sum of an $a^2$-improved topological
density definition after $\tau_\mathrm F=2.4 a^2$ units of Wilson flow,
as in \cite{Jahn:2020oqf}.  We find that values of
$N = 5$ or $N=10$ are adequate to preserve a good acceptance rate.

Lastly we remark on the optimal length of the individual
trajectories.  The figure of merit for trajectory length is the
mean-squared change in $S_U$ per unit numerical effort.  The numerical
effort is approximately linear in trajectory length $t_0$.  For a short
trajectory, $\Delta S_U$ is linear, and $(\Delta S_U)^2$ quadratic, in
trajectory length; but beyond a certain (fairly short) trajectory
length the action change saturates.  Therefore we started with a study
of how $(\Delta S_U)^2$ varies with trajectory length and chose the
value which maximizes $(\Delta S_U)^2/t_0$; $t_0 \simeq 0.75 a$.
A single trajectory leads to a change of
$(\Delta S_U / S_U)^2 \simeq 3/\Ndof$ where
$\Ndof = 24 V_L$ is the number of lattice
degrees of freedom (3 polarizations and 8 colors per site).
Therefore, for a well-chosen $W[S_U]$ function, since changes to $S_U$
accumulate in a Brownian fashion, the number of
updates needed to explore the full $\beta$ range is of order
\begin{equation}
  N_{\mathrm{updates}} \sim \Ndof \, \ln^2(\betamax/\betamin).
  \label{eq:Nupdates}
\end{equation}

\subsection{Choice and determination of $W[S]$}
\label{sec:bootstrap}

Now we return to the question of how to determine the weight function
$W[S]$.  We start by choosing the range of $\beta$ values we want to
explore, $\beta \in [\betamin,\betamax]$.  A short fixed-$\beta$
Markov chain establishes $\Smax = \langle S_U \rangle_{\betamin}$
and $\Smin = \langle S_U \rangle_{\betamax}$.
We then follow our procedure in \cite{Jahn:2018jvx,Jahn:2020oqf} and
choose a discrete set of values $S_i,i=(0,\ldots N_i)$ with
$S_0=\Smin$ and $S_{N_i} = \Smax$; $W[S_U]$ will be determined by its
values $W[S_i]$. However, because
the update described above works best when both $W[S_U]$ and $W'[S_U]$
are continuous functions, we shall interpolate $W[S_U]$ between these
points using a cubic spline function, rather than using a piecewise
linear function as in our previous papers.  We extend $W[S]$ above
$\Smax$ by choosing $W'[S>\Smax]=\betamin$ and similarly we set
$W'[S<\Smin] = \betamax$; these are also the values of the first
derivatives used in completing the definition of the spline function.

We will use the same automated improvement scheme to
determine $W[S_i]$ as we introduced in Ref.~\cite{Jahn:2018jvx}
(see also \cite{Laine:1998qk}).  We
generate a Markov chain using the update approach of the previous
subsection, but after each update, we adjust the
function $W[S_U]$ so as to make the current $S$ value less likely
(implying that the current $S$ is oversampled).  This is done by
\begin{enumerate}
\item
  determining $S_i,S_{i+1}$ such that the current $S$ value lies
  between them, $S_i < S_U < S_{i+1}$
\item
  increasing $W[S_i]$ and $W[S_{i+1}]$ by
  $\Delta W[S_i] = \frac{s_r (S_{i+1}-S_U)}{(S_{i+1}-S_i)}$
  and $\Delta W[S_{i+1}] = \frac{s_r(S_U-S_i)}{(S_{i+1}-S_i)}$.
  Here $s_r$ is an update strength which we explain next.
\item
  If $S_U$ is out of range, $W[S]$ is not updated.
\item
  If $i=0$ or $i+1=N_i$ so we are in the first or last interval,
  the boundary value is updated with double strength (since it
  is only updated half as often as other values).
\end{enumerate}
The value $s_r$ is initially chosen so that, in the time it takes for
the Monte-Carlo to go from the top to the bottom and back
(see \Eq{eq:Nupdates}), the $W[S_i]$ will change by of order 100.
Each time the $S$-value makes its way from the first interval to the
last and back (which we call a ``sweep''), we reduce $s_r$; at first
we reduce it by a factor of 2, but after the average $W[S_i]$ changes
by less than 1 per sweep, we change it by a reduced amount.  The update ends
when five sweeps change the average $W[S_i]$ by a total of less than 1.

\begin{figure}[t]
	\centering
	\includegraphics[scale=0.65]{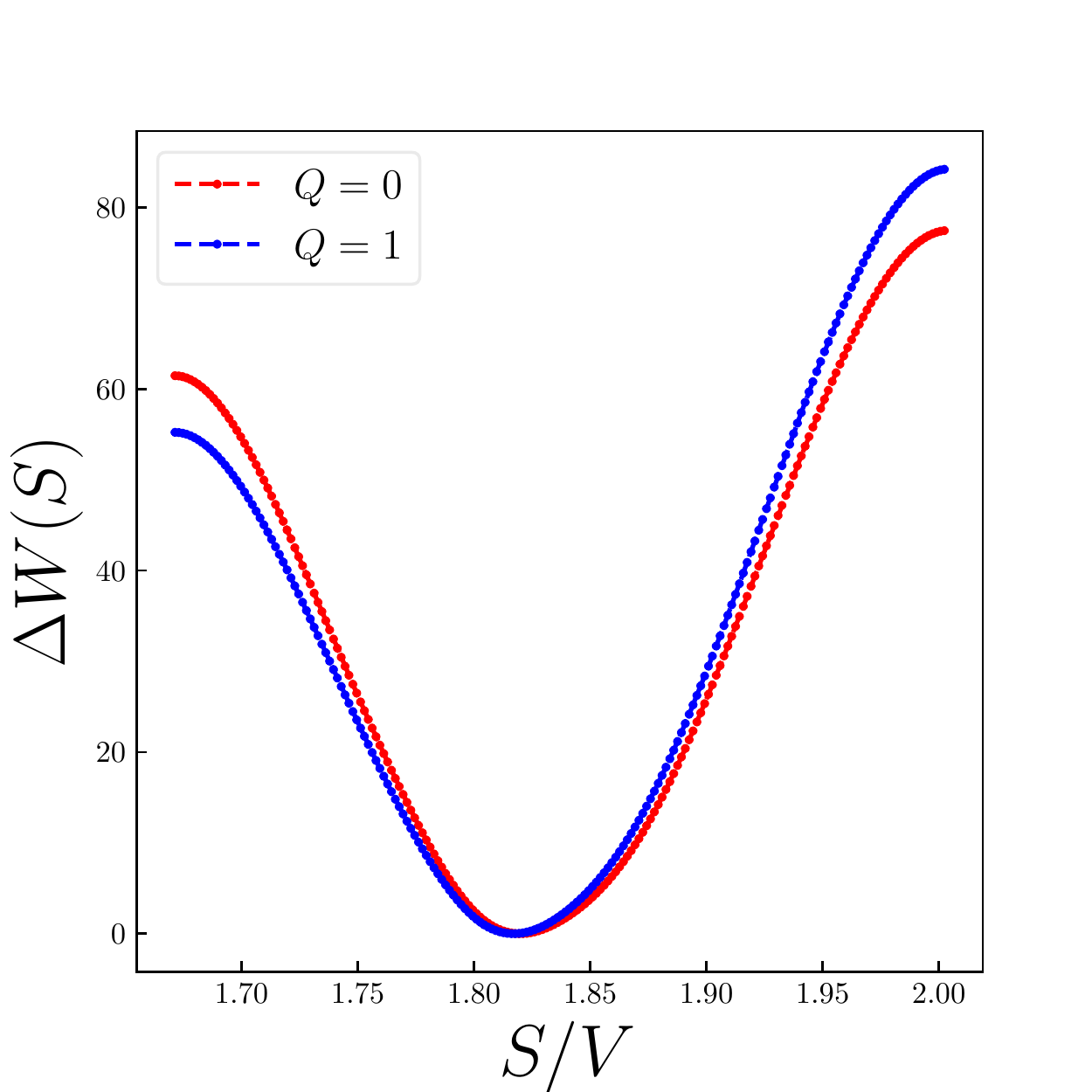}
	\caption{ Difference between the final determined $W[S]$ and the initial guess.}
	\label{fig:delta_W}
\end{figure}

\begin{figure}[t]
	\centering
	\includegraphics[scale=0.65]{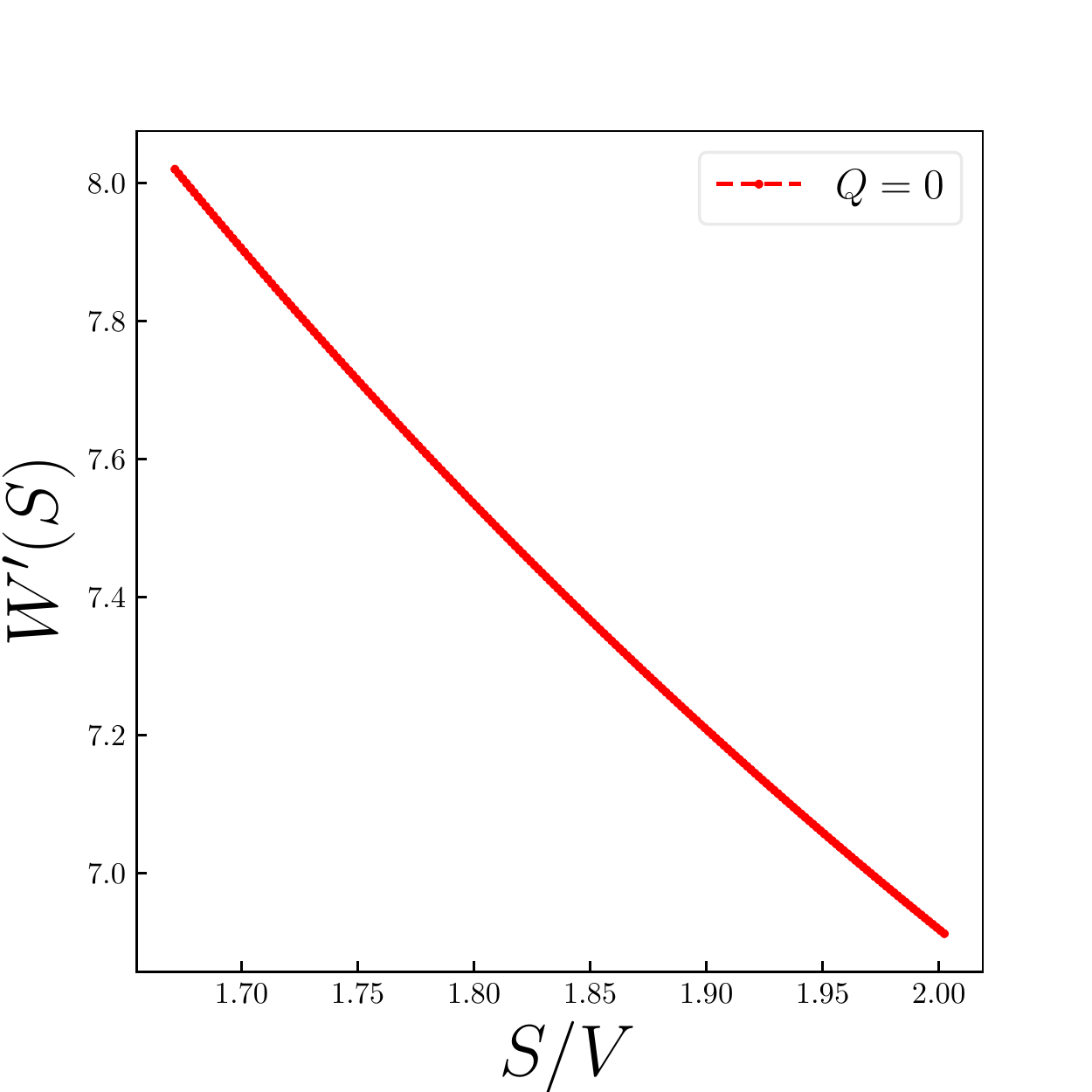}
	\caption{Final choice of $W'[S]$ for the lattice study
		described in Section \ref{sec:results}.}
	\label{fig:Wprime}
\end{figure}

This approach is inefficient if the initial $W$ function is very far
from its final form.  Therefore we improve the initial guess from the
simplest approach (that $W'[S]$ is linear).  Instead, we choose an
intermediate $\beta$ value $\betamid$ and perform a fixed-$\beta$
Monte-Carlo calculation to determine the associated $\Smid$ value.  We
then fit $W'$ to a quadratic, with values
$(\betamax,\betamid,\betamin)$ at the points $(\Smin,\Smid,\Smax)$,
integrate, and use this to determine starting guesses for the
$W[S_i]$.  With this approach, the initial and final $W$ functions
differ by less than 100, as shown in
 Figure \ref{fig:delta_W}, which displays the difference
between the initial and final $W$ function choice for the specific
lattice and $\beta$ range described in the next section.%
\footnote{We could further improve the initial guess
  by using more intermediate values;
  but using enough values, with precise enough $\langle S \rangle$
  determinations, to determine $W[S]$ to better than $\pm 1$
  costs the same as the $W$ refinement algorithm described here.}
The complete built $W^\prime$ is shown in Figure \ref{fig:Wprime}.

Finally, one must determine $W[S_U]$ for the $Q=1$ sector.  Here we
can take as an initial guess the $W[S_U]$ value determined in the
$Q=0$ sector.  To further refine this guess, we shift it by
\begin{equation}
  \label{Q1guess}
  W_{Q=1}[S] = W_{Q=0}[S] - 11 \ln(T/T_0)
\end{equation}
where $T$ is the temperature associated with the $\beta$ value
described by the slope $W'$ using a scale-setting relation between
lattice coupling $\beta$ and temperature $T$, and $T_0$ is a reference
temperature which could for instance be the temperature at
$\betamin$.  The factor 11 is the expected temperature dependence of
$\chitop a^4$ when the lattice spacing $a$ varies as $1/T$, at leading
perturbative order.  Again, \Eq{Q1guess} is only used to refine the
initial guess for $W_{Q=1}[S]$; we again perform an automated
$W$-changing  Markov chain to improve this guess; however the initial
$s_r$ value can be made much smaller.

After the $W[S_U]$-setting Markov chains are completed, we freeze the
values of $W_{Q=0,Q=1}[S]$ and use them in detailed-balance respecting
Markov chains which we will use to determine the susceptibility as
described in Subsection \ref{sec:suscept}.

\section{Results}
\label{sec:results}
\iffalse
\begin{table}[t]
	\centering
	\caption{Lattice parameters used in this work for building the $W(S)$ function.}
	\label{tab:ourlatts}
	\begin{tabular}{|c| c| c| c| c| c| c|}
		\hline \hline
		\multirow{2}{*}{$T \times L^2_{x,y} \times L_z$} & \multirow{2}{*}{$\beta_{\mathrm{min}}$} & \multirow{2}{*}{$\beta_{\mathrm{max}}$}&  \multicolumn{2}{c|}{\#Trajectories} & \multicolumn{2}{c|}{\# Sweeps}	\\ \cline{4-7}
		& & & $Q=0$ & $Q=1$ & $Q=0$ & $Q=1$ \\ 
		\hline
		$10 \times 32^2 \times 36$& 6.9076 & 8.01951 & $4 \times 10^6$ & $8 \times 10^6$& 22 & 17\\
		\hline\hline
	\end{tabular}
\end{table}
\fi
\begingroup
\renewcommand*{\arraystretch}{1.6}
\begin{table}[t]
	\centering
	\caption{\label{tab:ourlatts} Numerical cost of the simulation
          on a lattice four-volume $V = 10\times32^2\times36$ with
          $\betamin= 6.9076$ and $\betamax= 8.01951$.}
	\begin{tabular}{|c|c|c|c|}
		\hline
		Procedure & $Q$ & HMC Trajectories & Sweeps  \\ \hline 
		Building    &   0      &  $4 \times 10^6$ & 22 \\
		Building   &    1      &  $8 \times 10^6$ & 17 \\ \hline
		Measurements & 0 & $6.8 \times 10^6$ & 45  \\
		Measurements &  1 & $4.9 \times 10^6$ & 35 \\
		\hline
	\end{tabular}
\end{table}
\endgroup 
\begin{figure}[t]
	\label{fig:histogram}
	\includegraphics[scale=0.28]{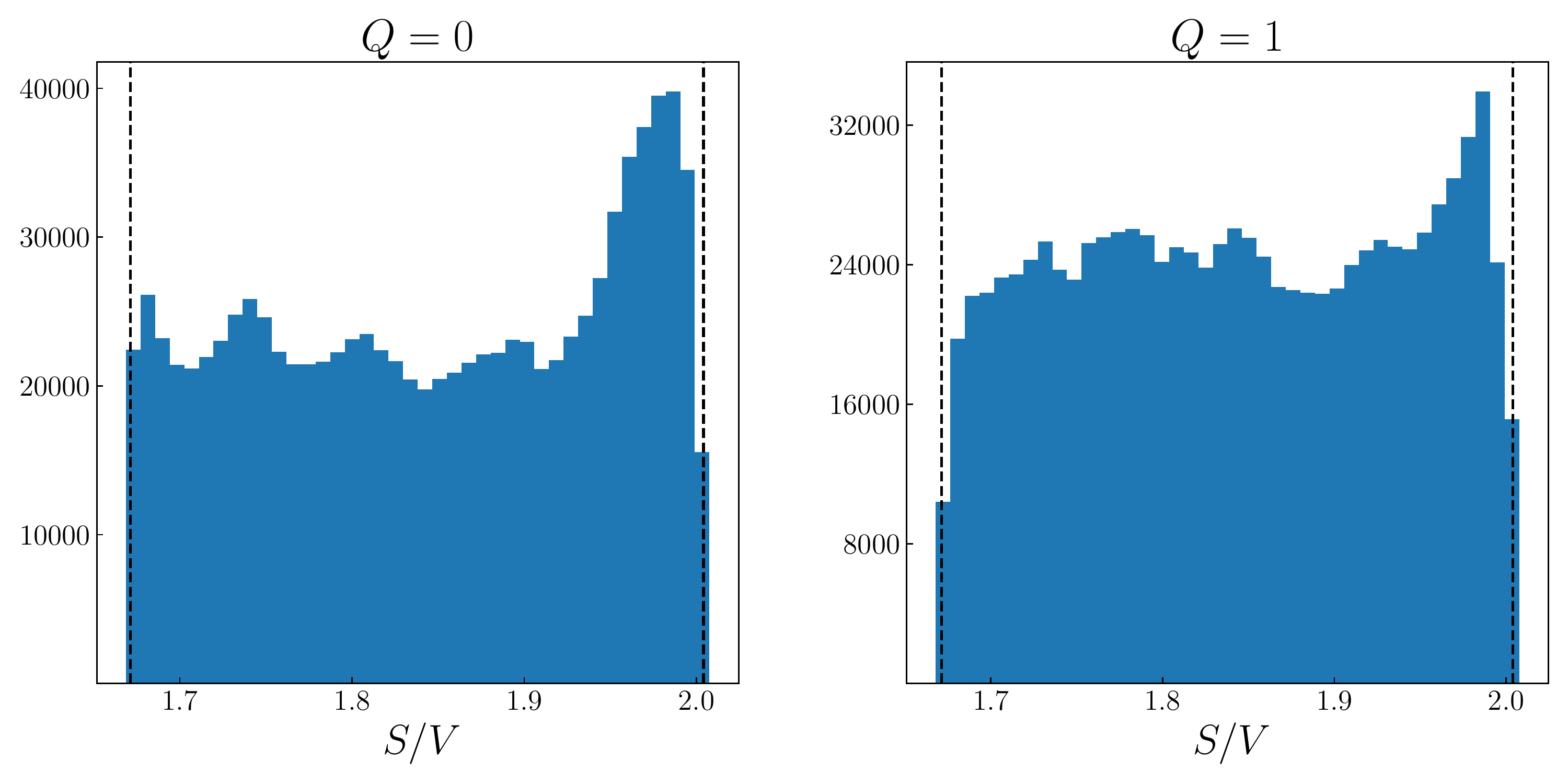}
	\caption{A histogram of the $S$ distribution of our Markov chain
		samples in the $Q=0$ sector (left) and the $Q=1$ sector
		(right).}
\end{figure}

In this section we present results of a simulation with aforementioned
$W(S)$ sampling.  Our goal is only to test the method; we will not
study multiple lattice spacings to attempt a continuum limit.
Instead, we choose one lattice geometry from among the geometries
studied in
Ref.~\cite{Jahn:2018dke,Jahn:2020oqf}, and which we can therefore use
to compare our results for susceptibility ratios to those obtained via
a competing technique.  Our goal is to establish which approach, the
one described here or the one described in Ref.~\cite{Jahn:2020oqf},
is more efficient at establishing the topological susceptibility at
high temperature.

We choose to investigate a lattice with temporal extent $N_\tau = 10$
and spatial extent $32^2 \times 36$, with
$(\betamin,\betamax)=(6.9076,8.01951)$,
which corresponds, according to the scale setting calculation of
Ref.~\cite{Burnier:2017bod} which we will use throughout, to
$T_{\betamin} = 2.5\, \Tc$ and $T_{\betamax} = 9.4 \, \Tc$.
This scale-setting relation involves applying a fit to scale-setting
data beyond the range where the reference has performed simulations,
that is, an extrapolation, which means it may not be absolutely
trustworthy.  However, by expressing our results in terms of
$\chitop(\beta) a^4(\beta)$, we can remain agnostic about the relation
between $T$ and $\beta$ and just determine how accurately we can
determine $\chitop a^4$ as a function of $\beta$
for \textsl{this specific} lattice geometry.  Getting continuum
results over a range of temperatures will then of course require
multiple lattice $N_\tau$ values and a reliable scale setting.
However, for the time being our goal is just to evaluate the
precision-to-numerical-cost ratio of the technique, so we leave this
problem for later.  The number of updates used in each procedure are
listed in Table \ref{tab:ourlatts}.  Note that an unfortunate choice
of $s_r$ made the $Q=1$ building unnecessarily inefficient; a more
careful choice should have required a fraction as many trajectories,
so that the trajectory count would be dominated by the measurements.

We begin with a check that our $W[S]$ function correctly generates a
rather uniform sample of configurations across the desired $\beta$
range.  We investigate this by plotting a histogram of the $S$ values
measured during the sampling Markov chain, both for the $Q=0$ and the
$Q=1$ ensembles, shown in Figure \ref{fig:histogram}.
The sample is adequately uniform.

\begin{table}[t]
  \centerline{ \begin{tabular}{|c|c|c|c|} \hline
     $T/\Tc$ & $\beta$ & $\ln(\chitop a^4(T)/\chitop a^4(2.8 \, \Tc))$
    & 1-$\sigma$ stat.\ err \\ \hline
    3.5 & $\;$7.1771$\;$ & $\phantom{1}$-2.25 & 0.17 \\
    4.0 & 7.2885 & $\phantom{1}$-3.68 & 0.20 \\
    5.0 & 7.4764 & $\phantom{1}$-5.79 & 0.24 \\
    7.0 & 7.7629 & $\phantom{1}$-9.39 & 0.30 \\
    9.0 & 7.9788 & -12.20 & 0.36 \\ \hline
  \end{tabular}}
  \label{FinalTable}
  \caption{Several temperatures, the corresponding $\beta$ values
    using the scale setting of Ref.~\cite{Burnier:2017bod}, and our
    results for the log susceptibility ratio and its 1-sigma statistical
    error.}
\end{table}

\begin{figure}[t]
	\centering
	\includegraphics[width=\linewidth]{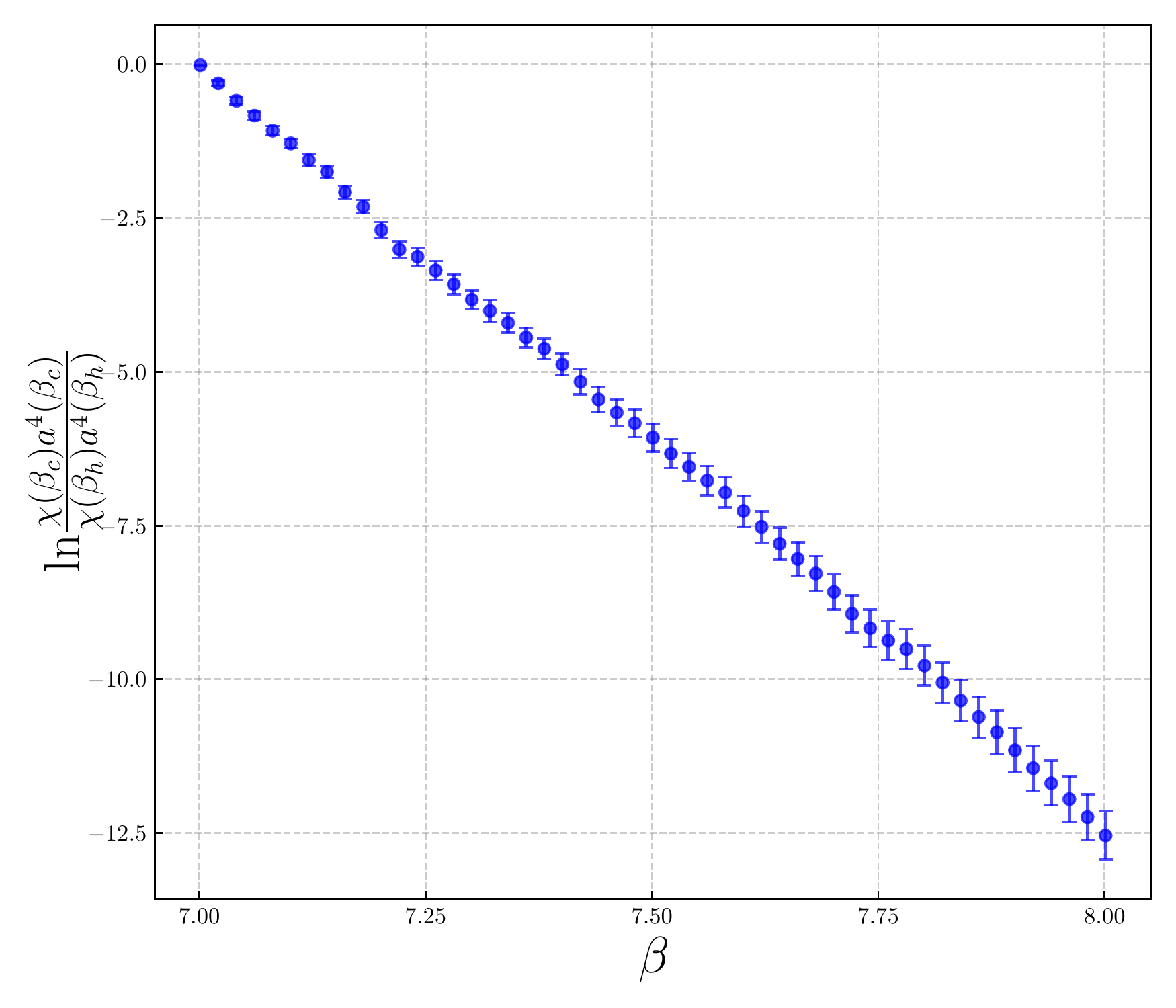}
	\caption{Results of the ratio in Eq.~\ref{ourway} }
	\label{fig:chi_ratio}
\end{figure}

Our main results are presented in Figure \ref{fig:chi_ratio}, which
shows how $\chitop a^4$ changes as a function of $\beta$ across the range
we study, evaluated from our Markov chains using \Eq{ourway}.  We have
chosen to use a value near the beginning of the $\beta$ range
($T=2.8 \, \Tc$) as the low temperature and to express all other
temperatures in relation to this one.  The figure, and further
data presented in Table \ref{FinalTable}, show that the error bars
are smallest between nearby $\beta$ values and grow to around
$\pm 0.35$ in $\ln(\chitop)$ for the widest-separated temperatures.
Our results agree within error bars of the results in
Ref.~\cite{Jahn:2020oqf} for those values where they are directly
comparable.

\section{Discussion}
\label{sec:discussion}
We have shown that the method we propose can successfully find the
$\beta$ dependence of $\chitop a^4$, and therefore the temperature
dependence of the susceptibility if the line of constant physics (that
is, $a(\beta)$) is known.  This determines $\chitop$ over a range of
temperatures if it is known at the lowest temperature, which is where
it is most easily determined by other approaches.

There are two key questions.  Is it more or less effective
than the rather similar approach of Refs
\cite{Borsanyi:2016ksw,Frison:2016vuc}?  And how does it compare with
the approach of Ref.~\cite{Jahn:2020oqf}?

The approach of Ref.~\cite{Borsanyi:2016ksw,Frison:2016vuc}
computes $\dd\ln(\chitop a^4)/\dd\beta$ at several $\beta$ values, which
it integrates to determine $\chitop a^4(\beta)$.  We review this approach
and compare it to our own in an appendix.  To summarize, in a
high-statistics determination, the approaches have essentially the
same numerical precision.  However, if lower precision is desired, the
numerical cost associated with building the $W(S)$ functions in our
approach is essentially ``dead weight'' which does not contribute to
the statistical power.  The other approach does not suffer from this
problem and so it is more efficient for a low-statistics
determination.  Our approach has the advantage that it automatically
includes all intermediate temperatures, while the alternative bases
the determined $\dd\ln(\chitop a^4)/\dd\beta$ on a finite set of values
which may leave discrete integration errors.  But it is not difficult
to use enough values to render this a minor concern.

Finally, we want to compare the numerical efficiency to the method of
Ref.~\cite{Jahn:2020oqf}.  Fortunately, the single lattice we
investigated in this work was also used in that reference, and we can
directly compare the error on $\chitop a^4(\beta_\mathrm h)/\chitop a^4(\beta_\mathrm c)$
found here with the error on the same quantity found there, along with
the number of HMC trajectories needed in each case.
In that reference $\chitop(T)$ was determined at
$\beta = (6.90097,7.30916,7.76294)$, corresponding to
$T = (2.5,4.1,7.0) \, \Tc$, a little narrower than the range considered here.  The
three determinations required a total of $9.2\times 10^6$
trajectories, about half the number which should have been needed
here.  The average trajectory length used was also shorter in that
reference than what we used here.  The the final errors on
$\ln(\chitop)$ in that study, for this lattice, were $(0.09,0.09,0.08)$
at the three temperatures.  In comparison, in comparing $2.8 \, \Tc$ to
$7.0 \, \Tc$ we find statistical errors of $0.30$.  To reduce these errors
to the level of the other study would therefore require about 10 times
more statistics in our measurement runs, indicating that the present
method is of order 10 times less numerically efficient.
Moreover, Ref.~\cite{Jahn:2020oqf} finds that the number of
trajectories needed for a given statistical error barely changes
as one increases the
volume (larger aspect ratio) or makes the lattice finer (larger
$N_\tau$ at fixed aspect ratio), whereas we know that the number of
updates needed for the method described in this paper should scale
with the number of lattice sites, see \Eq{eq:Nupdates}.

We conclude that our method is at least 10 times less
efficient than the single-temperature reweighting approach of
\cite{Jahn:2020oqf}, and will become still less efficient as one goes
closer to the large-volume and continuum limits.
As we understand it, this also implies that it should be easier in
principle to achieve small statistical errors with the approach
of \cite{Jahn:2020oqf} than with the approach of
\cite{Borsanyi:2016ksw,Frison:2016vuc}.

\begin{acknowledgments}
	The authors acknowledge support by the Deutsche Forschungsgemeinschaft
	(DFG, German Research Foundation) through the CRC-TR 211
	``Strong-interaction matter under extreme conditions'' -- project
	number 315477589 -- TRR 211. We also thank the GSI Helmholtzzentrum
	and the TU Darmstadt and its Institut f\"ur Kernphysik for supporting
	this research. Calculations were conducted on the Lichtenberg
	high performance computer of the TU Darmstadt. This work was performed
	using the framework of the publicly available openQCD-1.6 package
	\cite{openQCD}.
\end{acknowledgments}

\appendix
\section{Comparison with the slope method}

Our approach is closely related to the approach of
Refs.~\cite{Borsanyi:2016ksw,Frison:2016vuc}, and as we understand it,
the errors per numerical effort are nearly the same.  To explain this
conclusion, we start with a quick review of their approach, and then
look at the issue of statistical power in each approach.

Their approach also
seeks to compute \Eq{ratios} and then use a determination of
$\chitop(\beta_\mathrm c)$ to determine $\chitop$ at other temperatures.
Taking the log of \Eq{ratios} we find
\begin{align}
  \label{logs}
  \ln \frac{\chitop(\beta_\mathrm h) a^4(\beta_\mathrm h)}{\chitop(\beta_\mathrm c) a^4(\beta_\mathrm c)}
  & = \Big( \ln(Z_1(\beta_\mathrm h)) - \ln(Z_1(\beta_\mathrm c) \Big)
  \nonumber \\ & \phantom{=} {}
  - \Big( \ln(Z_0(\beta_\mathrm h)) - \ln(Z_0(\beta_\mathrm c) \Big) \,.
\end{align}
Now note that
\begin{align}
\label{derivatives}
  Z(\beta) & = \int \DD A_\mu e^{-\beta S}
  \quad \Rightarrow
  \nonumber \\
  - \frac{\partial \ln Z}{\partial \ln \beta}
  & = \frac{1}{Z} \int \DD A_\mu e^{-\beta S} \beta S
  = \langle \beta S \rangle
\end{align}
the $\beta$ dependence of $\ln Z$ is set by the expectation value of
the action.  Therefore
\begin{equation}
  \label{integrals}
  \ln Z_1(\beta_\mathrm h) - \ln Z_1(\beta_\mathrm c) =
  -\int_{\ln \beta_\mathrm c}^{\ln \beta_\mathrm h} \langle \beta S \rangle_1 \:
  \dd(\ln\beta)
\end{equation}
and the ratio we want is
\begin{equation}
  \label{theirway}
  \ln \frac{\chitop(\beta_\mathrm h) a^4(\beta_\mathrm h)}{\chitop(\beta_\mathrm c) a^4(\beta_\mathrm c)}
  = - \int_{\ln \beta_\mathrm c}^{\ln \beta_\mathrm h} \hspace{-0.7em}
  \left( \langle \beta S \rangle_1
  - \langle \beta S \rangle_0 \right) \dd(\ln\beta) \,.
\end{equation}
Refs.~\cite{Borsanyi:2016ksw,Frison:2016vuc}
evaluate $\langle \beta S \rangle$ in both $Q=0$ and $Q=1$
ensembles at a number of $\beta$ values, which are then used to
estimate this integral by, \textsl{eg}, the trapezoid rule.

To compute \Eq{theirway} using our approach, first write
\begin{align}
& \hspace{0.6em} \int_{\ln \beta_\mathrm c}^{\ln \beta_\mathrm h} -\langle \beta S \rangle \;
  \dd(\ln\beta)
  \nonumber \\
  & = \int_{\ln \beta_\mathrm c}^{\ln \beta_\mathrm h}
  \frac{-\sum_i e^{W[S_i] - \beta S_i} \beta S_i}
       {\sum_i e^{W[S_i]-\beta S_i}} \dd(\ln\beta)
  \nonumber \\
  & =  \int_{\ln \beta_\mathrm c}^{\ln \beta_\mathrm h}
  \frac{\dd}{\dd\ln\beta} \ln\left( \sum_i e^{W[S_i] - \beta S_i} \right)
  \dd(\ln\beta)
  \nonumber \\
  & = \ln \frac{\sum_i e^{W[S_i] - \beta_\mathrm h S_i}}
            {\sum_i e^{W[S_i] - \beta_\mathrm c S_i}}
\end{align}
so applying \Eq{expectation} to \Eq{theirway} leads directly to
\Eq{ourway}.  Therefore the real difference between the approaches is
whether \Eq{theirway} is estimated based on interpolating results for
several temperatures, or using a single Markov chain which spans all
temperatures.

Now consider the statistical power of each approach.  The accuracy of
a Monte-Carlo evaluation of $\langle \beta S \rangle$ is set by the
variance of $\beta S$ and the number of independent configurations
used.  The variance should be reasonably approximated as that for
$\Ndof$ Gau{\ss}ian random variables: $\sigma_{\beta S}^2 \simeq \Ndof/2$.
Therefore order-1 errors in $\langle \beta S \rangle$
require $\Ndof/2$ evaluations.  Since the
expectation value determines an integrand, this is multiplied by the
integration range, so $\Ndof/2$ evaluations return an error in the
ratio of partition functions which is of order $\ln(\betamax)-\ln(\betamin)$.
Evaluating $\langle \beta S \rangle$ at multiple $\beta$ values leads
to a larger error at each evaluation, but because each is responsible
for a narrower $\Delta \beta$ range and the errors are uncorrelated,
the final statistical uncertainty is independent of the number of
$\beta$ values used in the evaluation and depends only on the total
number of Markov steps and the width of the $\beta$ range considered.
The final error estimate is
$\Delta \ln\chi = \ln(\betamax/\betamin)\,
\sqrt{\Ndof/2N_{\mathrm{updates}}}$.  The error rises by $\sqrt{2}$
and $N_{\mathrm{updates}}$ is doubled when we recall that separate
simulations are needed in the $Q=0$ and $Q=1$ sectors.

In comparison, we see in \Eq{eq:Nupdates} that our approach can
explore the full $\beta$ range, leading to order-1 errors in
$\ln\chi$, in $N_{\mathrm{updates}} \sim \Ndof \,\ln^2(\betamax/\betamin)$.
Therefore the two approaches produce
errors per unit numerical effort which are the same up to an
order-1 factor.  In a numerical experiment on a toy problem
($N$ independent Gau{\ss}ian random variables $x$ with action
$S=\sum x^2/2$) we find that the order-1 factor is in fact 1, so
the two approaches have the same statistical power per compute time,
provided that $W[S]$ is well determined and neglecting the
computational effort expended in evaluating it.

We should also remark on how each approach is extended to
full (unquenched) QCD.  In each case the main challenge is dealing
with the way quark masses must be varied with the lattice spacing
and therefore with $\beta$: $m=m(\beta)$ (which must also be
determined as part of the scale setting).
This added $\beta$ dependence changes \Eq{derivatives}, replacing
$\langle \beta S \rangle \to \langle \beta S
+ \beta \, \dd m/\dd\beta \, \bar\psi\psi \rangle$.
In our approach one must replace
$W[S] \to W[S] + \bar\psi (\slashed{D}+m(W'[S])) \psi$
where we use $W'$ in place of $\beta$ for the scale dependence of $m$.
This amends \Eq{eomP} by the addition of the standard fermionic
force term and by the replacement
$\dd W/\dd S_U \to (\dd W/\dd S_U)
+ (\dd m/\dd\beta)(\dd W'/\dd S_U)\langle \bar\psi\psi \rangle$
where $\langle \bar\psi\psi \rangle$ is the sum of the $\bar\psi\psi$
value over all
sites in the current configuration.  Finally, in \Eq{ourway}, the
$W-\beta S$ reweighting must be complemented by a determinant-ratio
from the $S$-dependent mass to the physical mass for the desired
$\beta$ value.

\bibliographystyle{apsrev4-1}
\bibliography{refs}
\end{document}